\documentclass[a4paper]{jpconf}
\usepackage{graphics}
\usepackage{graphicx}
\usepackage{amsmath}

\newcommand{\be}{\begin{equation}}
\newcommand{\ee}[1]{\label{#1} \end{equation}}
\newcommand{\ba}{\begin{eqnarray}}
\newcommand{\ea}[1]{\label{#1} \end{eqnarray}}
\newcommand{\nl}{\nonumber \\}

\newcommand{\pd}[2]{ \frac{\partial {#1}}{\partial {#2}}  }
\newcommand{\re}[1]{(\ref{#1})}

\begin{document}

\title{Nonadditive thermostatistics and thermodynamics}

\author{ 
  P V\'an\(^{1,2}\), G G Barnaf\" oldi$^1$, T  S Bir\'o$^1$,\ and K  \"Urm\" ossy\(^1\) 
}
\address{ \(^{1}\) Wigner RCP, Hungarian Academy of Sciences
  { H-1525 Budapest, P.O.Box 49, Hungary} }
\address{\(^{2}\)Department of Energy Engineering, Budapest University of Technology and Economics,
  Bertalan Lajos u. 4-6, H-1111, Budapest, Hungary}
\ead{van.peter@wigner.mta.hu}

\begin{abstract}
Nonadditive composition rules for several physical quantities are treated in  thermodynamics. It is argued that the zeroth law defines the existence of their additive forms, the formal logarithms. A further principle, the universal thermostat independence leads to a particular formal logarithm, equivalent to Tsallis entropy $S_q$. We connect $q$ with generalized susceptibilities of the thermostat.
\end{abstract}

\section{Introduction}

Neither thermodynamics nor statistical physics includes the other. Near-equilibrium statistical physics fulfills thermodynamic relations particular microscopic models can clarify certain mechanisms behind thermodynamic quantities. On the other hand the general laws of thermodynamics are valid also in those cases when the detailed microscopic treatment is unavailable or hopelessly complicated. The aiming at universality in thermodynamics should be understood from this point of view. Whenever one deduces model relations from thermodynamic principles, the results are by construction independent of particularities of the microscopic model.

\begin{figure}
\begin{center}
\includegraphics[width=0.66\textwidth,angle=0]{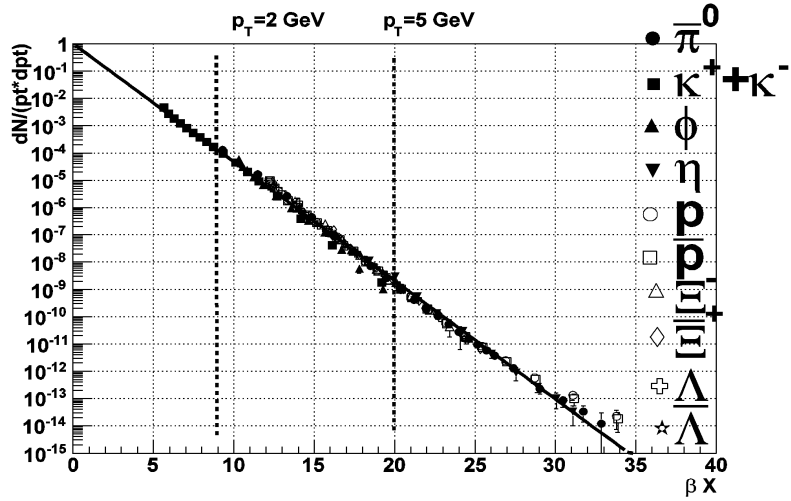}
\end{center}
\caption{ \label{hi}
Tsallis fits to transverse momentum distributions of identified particles stemming from AuAu collisions at $\sqrt{s} =$ 200 AGeV collision energy. The expansion of the quark-gluon plasma is taken into account by the blast wave model. X = $\frac{1}{a} ln(1 + a\gamma(m_T - vp_T )$) and $\beta=1/T$.  (Fig. 1b in Ref. \cite{BirEta10a}). 
}
\end{figure}

\begin{figure}
\begin{center}
\includegraphics[width=0.66\textwidth,angle=0]{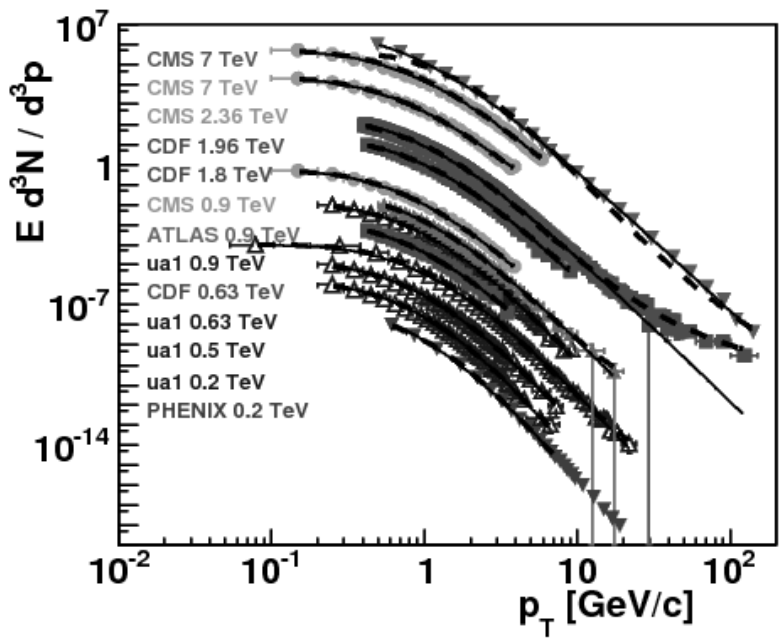}
\end{center}
\caption{ \label{pp}
   Tsallis fits on transverse momentum spectra of charged hadrons in proton-proton collision on a logarithmic plot (Fig. 1a in Ref. \cite{BarEta11a}). 
}
\end{figure}

\begin{figure}
\begin{center}
\includegraphics[width=0.66\textwidth,angle=0]{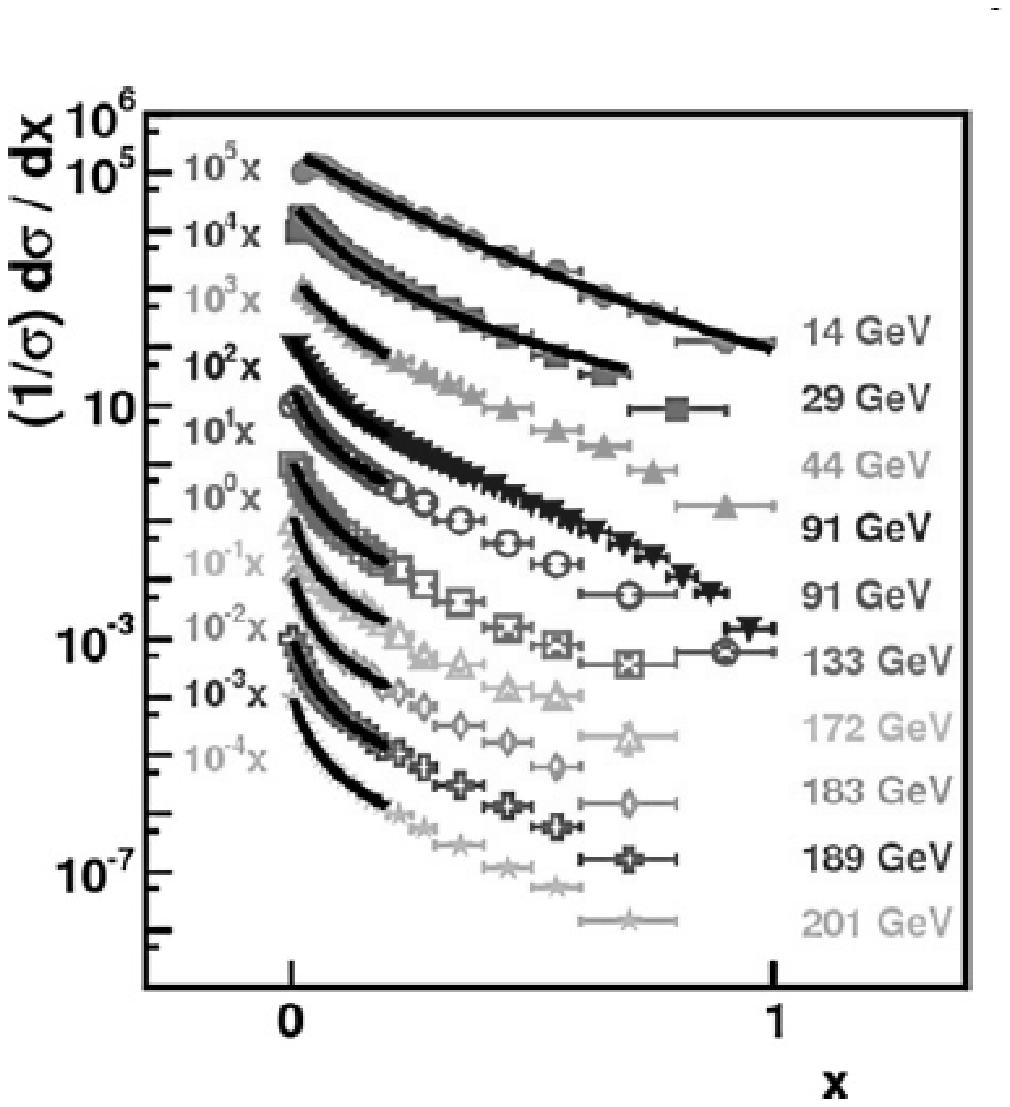}
\end{center}
\caption{ \label{ee+}
Fragmentation functions from electron-positron collisions measured at various collision energies and fitted one-dimensional Tsallis distributions. The energy fraction x = $\epsilon/E$ is the energy of the produced hadron scaled by the maximal acquirable energy in a 2-jet event E = $\sqrt{s}/2$ (Fig. 5 in Ref. \cite{UrmEta11a}). 
}
\end{figure}

The above relation is present in attempts of explaining among others power-law tails in the transverse momentum spectra of high energy collisions based on statistical considerations. These power-law tails can be observed in heavy ion collisions at the RHIC and LHC accelerators \cite{BirEta10a} cf. Fig. \ref{hi}, in proton-proton collisions at the LHC, RHIC, Fermilab and SPS \cite{BarEta11a} cf. Fig. \ref{pp} and even in electron-positron collisions \cite{UrmEta11a}, cf. Fig. \ref{ee+}. One inspects, that in electron-positron collisions, where the number of outcoming particles is about 10-50, the deviation from Tsallis distribution is salient. 


The presented data indicate the universal nature of the cut power law distributions. These fits perform much better also at low $p_T$ than the exponential Boltzmann-Gibbs distribution \cite{CMS10a}. Therefore a single microscopic mechanism,  like perturbative QCD, cannot explain all observations. 

In this paper we propose a universal explanation of the observed Tsallis distributions. This is based on general arguments extending the framework of classical thermostatics by abandoning the additivity property constraint on the extensive quantities by introducing abstract, in a general setting nonadditive  composition rules \cite{Abe02a,Bir08a}. By doing so the zeroth law of thermodynamics reveals that the very concept of separate physical systems -- thermodynamic bodies -- restrict those rules. In the following we are going to prove that rules fulfilling the zeroth law are associative and can be given in a special form with the help of the so called formal logarithms. Then the thermodynamic temperature can be introduced as an intensive quantity characterizing the thermodynamic bodies separately. The extension of the maximum entropy principle for nonadditive composition rules becomes straightforward. Nonadditively extended thermodynamics ensures that thermostatistical relations -- represented by different 
statistical 
entropy 
formulas -- can be interpreted in a thermodynamic framework \cite{BirVan11a}. 

This framework can be further refined when dealing with the choice among several different generalizations of thermostatistical entropies (e.g. \cite{Tsa88a,Tsa09b,Nau03a,Arn01b,Tan01b,Kan01a,LanVed98a}. One would welcome a reasonable criteria formulated in general thermodynamics \cite{BirEta12m}. We show that considering a canonical, maximum entropy based approach to an open statistical physical system. The requirement of vanishing first order corrections in the microcanonical treatment of a system including the thermostat, leads to a particular formal logarithm, that transforms the Boltzmann\,--\,Gibbs\,--\,Shannon entropy density to a Tsallis entropy density. This leads to a physical interpretation of the {\em q} parameter in the entropy formula, related to the `heat capacity' of the reservoir. In this way the Tsallis entropy density and the R\'enyi total entropy are connected by the selected formal logarithm.

\section{The zeroth law of thermodynamics and nonadditive composition rules}

\subsection{Separability}

The basic variables of classical thermodynamics are the so called extensive physical quantities, like the energy ($E$), particle number of the i\textsuperscript{th} chemical component ($N_i$)  and entropy ($S$). The adjective 'extensive' is often associated to a different characteristics of these fundamental physical quantities. It is customary to assume that the the extensive quantities are additive if we merge two thermodynamic bodies:
\begin{equation}
        X_{12}=X_{1}+X_{2},
\end{equation}
where $X_i$ denotes one of the above mentioned quantities of the {i}\textsuperscript{th} body. This special property of composition is \textit{additivity}. On the other hand traditionally one assumes that these quantities characterize the systems up to the largest meaningful scale operating with finite densities of the extensive quantities \cite{Mac92b}:
\be
 \rho_X = \lim_{N\rightarrow\infty} \frac{1}{N}\sum_{i=1}^N X_i < \infty.
\ee{ext}  
Here one divides the body to \textit{N} different parts and \(X_i\) belongs to the  i\textsuperscript{th} of them. This property is the \textit{extensivity}. These two properties are related, but far from being equivalent. If a quantity is additive, then it is necessarily extensive, but there are  extensive and yet non-additive quantities \cite{Bir08a,Tsa09b}. In our treatment the classically extensive quantities are in general nonadditively composed, therefore in the followings we refer to them as {\em basic quantities}.

Another important class of thermodynamic quantities are the intensives. Their definition is related to thermodynamic equilibrium and to the zeroth and the second laws of thermodynamics. The zeroth law is  formulated as the requirement of the \textit{transitivity} of the thermodynamic equilibrium state \cite{Max902b,FowGug39b}. Somewhat implicit, but important in this respect, that the thermodynamic bodies are assumed to be separable of each other: in thermodynamic equilibrium their respective equation of states  do not depend on the properties of the partner bodies or of the interaction. This {\em condition of physical separability} is required also in case of nonadditive composition of the extensives. We shall introduce the concept of intensives accordingly. In the following we generalize our reasoning with a nonadditive entropy and nonadditive energy as basic quantities developed in \cite{BirVan11a} to nonadditive entropy and additional $n$ nonadditive quantities. 

Let us consider two thermodynamic bodies, characterized by {\em n} basic quantities denoted by $X_1^i$ and $X_2^i$, $i=1,...,n$ for the two bodies respectively and $X^0_1=S_1$, $X^0_2=S_2$ are the entropies. The entropy plays a distinguished role, it  defines the equation of state of the respective bodies, $S(X^1,...,X^n)$. In our treatment the basic quantities are not additive, but composable, therefore their value in the two-body system is expressed by the composition rules 
\ba
X^i_{12}(X_1^i,X_2^i) \neq X^i_1 + X^i_2, \qquad i=0,...,n.
\ea{comprule}

This functional relationship is in general nonadditive. We assume that our thermodynamic system of two interacting bodies is isolated, therefore the basic quantities, $X_{12}^i$ are fixed. Their conservation is expressed with the help of differentials:
\ba
\text{d}X_{12}^i = \pd{X^i_{12}}{X^i_1}\text{d}^i_1+\pd{X^i_{12}}{X^i_2}\text{d}X^i_2=0, \qquad i=1,...,n.
\ea{consX}
 
The requirement of maximum entropy in isolated systems in equilibrium reads as
\ba
\text{d}S_{12}(S_1,S_2) = \sum_{i=1}^n \left(\pd{S_{12}}{S_1}\pd{S_1}{X^i_1}\text{d}X^i_1+\pd{S_{12}}{S_2}\pd{S_2}{X^i_2}\text{d}X^i_2\right) = 0.
\ea{consS}

The unique condition of the equations \re{consX} and \re{consS} having arbitrary $\text{d}X_1^i$ and $\text{d}X^i_2$ changes in the basic quantities is given by
\ba
\pd{X_{12}^i}{X_2^i}\pd{S_{12}}{S_1}\pd{S_1}{X^i_1} = \pd{X_{12}^i}{X_1^i}\pd{S_{12}}{S_2}\pd{S_2}{X^i_2}, \quad i=1,...,n.
\ea{eqcond}

The most general form of the partial derivatives now may be written by a multiplicative separation of the variables as 
\ba
 \pd{S_{12}}{S_1}(S_1,S_2) &=& F_1(S_1)\ G_2(S_2)\ H_1(S_1,S_2),
\nl
 \pd{S_{12}}{S_2}(S_1,S_2) &=& F_2(S_2)\ G_1(S_1)\ H_2(S_1,S_2),
\nl
 \pd{X^i_{12}}{X^i_1}(X_1^i,X_2^i) &=& A^i_1(X^i_1)\ B^i_2(X^i_2)\ C^i_1(X^i_1,X^i_2),
\nl
 \pd{X^i_{12}}{X^i_2}(X_1^i,X_2^i) &=& A^i_2(X^i_2)\ B^i_1(X^i_1)\ C^i_2(X^i_1,X^i_2).
\ea{parcd_fact}

Then \re{eqcond} requires, that
\be
 A^i_2B^i_1C^i_2 \cdot F_1G_2H_1 \pd{S_1}{X^i_1} = A^i_1B^i_2C^i_1 \cdot F_2G_1H_2 \pd{S_2}{X^i_2}.
\ee{facteq}
This equation factorizes to $(X^i_1,S_1)$ and $(X^i_2,S_2)$ dependent terms only if
\be
\frac{C^i_2(X^i_1,X^i_2)}{C^i_1(X^i_1,X^i_2)} =  \frac{H_2(S_1,S_2)}{H_1(S_1,S_2)}
\quad \text{for}\quad i=1,...,n
\ee{gfact_cond}
The physical separability of the thermodynamic bodies has to beindependent of any particular form of the equation of state, $S(X^1,...,X^n)$. Therefore the above ratio can only be constant. Its value can easily be absorbed into one of the factorizing component functions of the entropy $F$ or $G$. As an immediate consequence one obtains
\ba
 C^i_1(X^i_1,X^i_2) = C^i_2(X^i_1,X^i_2),
\nl
 H_1(S_1,S_2) = H_2(S_1,S_2).
\ea{sepcond}
These equalities are the basis for considering formal logarithms for the basic quantities separately. The factorized form of eq. \re{eqcond}
\be
\frac{B^i_1 F_1}{A^i_1 G_1} \pd{S_1}{X^i_1} = \frac{B^i_2 F_2}{A^i_2 G_2} \pd{S_2}{X^i_2},
\ee{0lawfact}
defines the following generalized intensive quantities $Y^i$
\be
 Y^i = \frac{B^i(X^i)F(S)}{A^i(X^i)G(S)} \pd{S}{X^i}(X^1,...X^n).
\ee{genint1}
Finally, using the definitions
\ba
 \hat{L}(S) := \int \frac{F(S)}{G(S)} dS,
\nl
 L^i(X^i) := \int \frac{A^i(X^i)}{B^i(X^i)} dX^i,
\ea{formlogdefs}
we arrive at
\be
 Y^i = \pd{\hat{L}(S)}{L^i(X^i)}.
\ee{gentemp2}
The zeroth law requires that this common value is to be introduced as the generalized entropic intensive quantity. 

These functions of the original thermodynamical variables, $\hat{L}(S)$ and $L^i(X^i)$
defined in eq.(\ref{formlogdefs}) can be used to map the original composition
rules to a simple addition. 

Namely considering the conditions \re{sepcond} one obtains
\ba
\frac{B^i_1}{A^i_1} \pd{X^i_{12}}{X^i_1} &=& \frac{B^i_2}{A^i_2} \pd{X^i_{12}}{X^i_2},
\nl
\frac{G_1}{F_1} \pd{S_{12}}{S_1} &=& \frac{G_2}{F_2} \pd{S_{12}}{S_2}.
\ea{compositaddons}
Utilizing now the definitions (\ref{formlogdefs})  for $X^i_1,X^i_2$, $i=1,...,n$ and 
and $S_1,S_2$ separately, the partial derivatives simplify:
\ba
  \pd{X^i_{12}}{L^i_1} &=& \pd{X^i_{12}}{L^i_2},
\nl
\pd{S_{12}}{\hat{L}_1} &=& \pd{S_{12}}{\hat{L}_2}.
\ea{formderiv}
The general solution of such partial differential equations is an arbitrary function of
the sum of variables:
\ba
 X^i_{12} &=& \Phi^i(L^i_1+L^i_2), \quad \text{for all} \quad i=1,...n.
\nl
S_{12} &=& \Psi(\hat{L}_1+\hat{L}_2).
\ea{sum}
If the $\Phi$ and $\Psi$ functions are invertible we can index their inverse arriving at a more symmetric notation:
\ba
L^i_{12}(X^i_{12}) &=& L^i_1(X^i_1) + L^i_2(X^i_2),
\nl
\hat{L}_{12}(S_{12}) &=& \hat{L}_1(S_1) + \hat{L}_2(S_2).
\ea{mapaddrules}
Since the $L^i$ and $\hat{L}$ functions map a non-additive composition rule to an additive one, they are {\em formal logarithms}.

One recovers the classical additive composition when the formal logarithms are the respective identity functions. The role of the zeroth law in abstract composition rules was investigated by several authors with different generality and purpose  \cite{Abe01a,Abe01a1,Abe02a,Joh03a,Joh04a,Sca10a} and with diverging conclusions.  

Composition rules, that can be transformed into the above form with a formal logarithm, are associative. Associative composition rules are also results of an infinite repetitive application of the same rule onto infinitesimal pieces of the same material, as it was proved in \cite{Bir08a}. However, in this special case the formal logarithms in \re{mapaddrules} are related to the same material and therefore are identical. In the general, heterogeneous  case these functions $L_1$ and $L_2$ may differ, and the mapping related to the interaction $L_{12}$ can be a further one.

\subsection{Transitivity}

Up to now we required the separability of the thermodynamic bodies in equilibrium and did not investigate the transitivity aspect of the zeroth law applying nonadditive composition rules. This formulation is not connected to  the second law directly. Transitivity universally requires that if bodies 1 and 2 are in thermal equilibrium and independently the bodies 2 and 3, then also the bodies 1 and 3 are in equilibrium \footnote{J. C. Maxwell was the first who formulated the zeroth law in this form \cite{Max902b}.  A representative summary of the history is given here: \textit{http://www.eoht.info/page/Zeroth+law+of+thermodynamics}}. 

So far we have established additivity of composite functions of the basic quantities of the respective subsystems, $L_I^i(E_I^i)$, $I=1,2 $. Therefore it is natural to assume that these functions are characteristic to the bodies and only the double-indexed formal logarithms, $L^i_{IJ}(X^i_{IJ})$, $I,J=1,2$ are characteristic to the interaction between bodies in equilibrium. This is valid for the entropy composition, as well. By this construction all subsystems develop the same individual formal logarithm, irrespective to which other system they equilibrate with. Assuming namely the opposite, i.e. a partner-dependent individual formal logarithm, the transitivity would be violated. Let us regard three possible pairings of three subsystems. The composite basic quantities satisfy
\ba
X^i_{12} &=& \Phi^i_{12}(L^i_1(X^i_1)+L^i_2(X^i_2)),
\nl
 X^i_{23} &=& \Phi^i_{23}(\tilde{L}^i_2(X^i_2)+L^i_3(X^i_3)),
\nl
 X^i_{13} &=& \Phi^i_{13}(\tilde{L}^i_1(X^i_1)+\tilde{L}^i_3(áX^i_3)).
\ea{three_composit}
If $\tilde{L}^i \ne L^i$, then the equilibrium condition is not automatically transitive. The same is valid for the nonadditive entropy, too:
\ba
S_{12} &=& \Psi_{12}\left(\hat L_1(S_1)+\hat L_2(S_2)\right),
\nl
S_{23} &=& \Psi_{23}\left(\check{L}_2(S_2)+\hat L_3(S_3)\right),
\nl
S_{13} &=& \Psi_{13}\left(\check{L}_1(S_1)+\check{L}_3(S_3)\right).
\ea{Sthree_composit}

Equation (\ref{eqcond}) requires, that  
\ba
 \pd{S_{12}}{S_1} \pd{S_1}{X^i_1} \pd{X^i_{12}}{X^i_2} &=& 
   \pd{S_{12}}{S_2} \pd{S_2}{X^i_2} \pd{X^i_{12}}{X^i_1}
\nl
 \pd{S_{23}}{S_2} \pd{S_2}{X^i_2} \pd{X^i_{23}}{X^i_3} &=& 
   \pd{S_{23}}{S_3} \pd{S_3}{X^i_3} \pd{X^i_{23}}{X^i_2}
\nl
 \pd{S_{13}}{S_1} \pd{S_1}{X^i_1} \pd{X^i_{13}}{X^i_3} &=& 
   \pd{S_{13}}{S_3} \pd{S_3}{X^i_3} \pd{X^i_{13}}{X^i_1}
\ea{transitiv_zero_cond}
From here the following separation is obtained, also considering \re{mapaddrules}: 
\ba
  \frac{\pd{S_{12}}{S_1} \pd{S_1}{X^i_1}}{\pd{S_{12}}{S_2} \pd{S_2}{X^i_2}} &=& 
    \frac{\pd{X^i_{12}}{X^i_1}}{\pd{X^i_{12}}{X^i_2}} = 
    \frac{\hat L_1^{\prime}S_1^{\prime}}{\hat L_2^{\prime}S_2^{\prime}} =
    \frac{L^{i\prime}_1}{L^{i\prime}_2},
\nl 
  \frac{\pd{S_{23}}{S_2} \pd{S_2}{X^i_2}}{\pd{S_{23}}{S_3}\pd{S_3}{X^i_3}} &=& 
    \frac{\pd{X^i_{23}}{X^i_2}}{\pd{X^i_{23}}{X^i_3}} = 
    \frac{\check{L}_2^{\prime}S_2^{\prime}}{\hat L_3^{\prime}S_3^{\prime}} =
    \frac{\tilde L^{i\prime}_2}{L^{i\prime}_3},
\nl 
  \frac{\pd{S_{13}}{S_1} \pd{S_1}{X^i_1}}{\pd{S_{13}}{S_3} \pd{S_3}{X^i_3}} &=& 
    \frac{\pd{X^i_{13}}{X^i_1}}{\pd{X^i_{13}}{X^i_3}} = 
    \frac{\check{L}_1^{\prime}S_1^{\prime}}{\check{L}_3^{\prime}S_3^{\prime}} =
    \frac{\tilde L^{i\prime}_1}{\tilde L^{i\prime}_3}.
\ea{three_ratios}
This condition can be reduced easily to obtain
\be
  \frac{\hat L^{\prime}_1\check{L}^{\prime}_2\check{L}^{\prime}_3}
    {\hat{L}^{\prime}_2\hat L^{\prime}_3\check{L}^{\prime}_1} =
  \frac{L^{i\prime}_1\tilde{L}^{i\prime}_2\tilde{L}^{i\prime}_3}
    {{L}^{i\prime}_2L^{i\prime}_3\tilde{L}^{i\prime}_1} 
\ee{three_reduced}
for $i=1,...,n$. The left hand side of the $n$ equalities depends on the body entropies $S_1$, $S_2$ and $S_3$ and the right hand side depends on the $i$\textsuperscript{th} basic quantities of the bodies $X^i_1$, $X^i_2$ and $X^i_3$, respectively. Moreover it is valid for any permuted arrangement of the lowercase body indices $1$, $2$ and $3$. Therefore one concludes that the transitivity of thermal equilibrium can be satisfied only if the 
\be
\frac{\hat L^{\prime}(S)}{\check{L}^{\prime}(S)} = const. \quad \text{and} \quad
\frac{L^{i\prime}(X^i)}{\tilde{L}^{i\prime}(X^i)} = const. 
\ee{transconcl} 

This is a necessary and sufficient condition. Physically sensible composition rules must satisfy two additional simple requirements. The \textit{triviality condition} says, that a composition with zero does not change the value, therefore $L(0)=0$ is also required. According to the \textit{compatibility condition}  for small values of the basic quantities the nonadditive effects are reduced and addition emerges, therefore $L^{\prime}(0)=1$. Due to these two conditions, from eq. \re{transconcl} it follows that the  $\tilde{L}^i$ and $L^i$ functions of the basic quantities are identical, as well as the $\hat L$ and $\check L$ functions of the entropy. 

One aspect of the zeroth law -- the separability -- enforces that composition rules are expressed by formal logarithms. Another aspect of the zeroth law -- the transitivity -- ensures that $\hat{L}_I$ and $L^i_I$ are characteristics of the thermodynamic body $I$ and only $L_{IJ}$ and $\hat{L}_{IJ}$ depend on the interactions. It is straightforward to see, that zeroth law compatible composition rules are associative if they are \emph{homogeneous}, that is $L_{IJ}=L_I=L_J$.

\subsection{First order compositions}

A particular non-additive entropy formula, advanced by Tsallis, underlies the following composition rule \cite{Tsa09b,Tsa88a}
\be
S_{12}(S_1,S_2) = S_1 + S_2 +{a} S_1 S_2.
\ee{TSALLIS_COMBO}
We have introduced the shorthand notation $\hat{a}=1-q$. Additive entropy systems
realize ${a}=0$, non-additive systems a non-zero value of this parameter. One can get the above  formula as a Taylor series expansion of an arbitrary $S_{12}(S_1,S_2)$ up to first order in $S_1$ and $S_2$ and requiering the homogeneity of the composition.  

According to our previous result  a thermodynamic framework requires formal logarithms. The formal logarithm for the above rule is easy to derive from
\be
1 + {a} S_{12} = 1 + {a}S_1 + {a}S_2 + {a}^2 S_1 S_2 
        = (1+{a}S_1) \, (1+{a}S_2).
\ee{TSALLIS_EASY}
The product is related to the addition by the logarithm and scaled down to satisfy
$\hat{L}^{\prime}(0)=1$:
\be
\hat{L}(S) = \frac{1}{{a}} \ln (1+{a}S).
\ee{TSALLIS_FORMAL_LOG}

From the viewpoint of zeroth law compatibility it is strongly advised to consider the additive formal logarithm. In case of the Tsallis entropy,
\be
S_T = \frac{1}{{a}} \left( \sum_i\limits p_i^{1-\hat{a}} - p_i \right),
\ee{TSALLIS_FORMULA}
with $\hat{a}=1-q$ and the normalization $\sum_i p_i = 1$, its formal logarithm is given by the well known R\'enyi entropy \cite{Ren61a}:
\be
S_R = \hat{L}(S_T) = \frac{1}{\hat{a}} \ln (1+\hat{a}S_T) = \frac{1}{1-q} \ln \sum_i\limits p_i^{q}.
\ee{RENYI_FORMULA}

\section{Universal Thermostat Independence}

The analysis of the previous section shows the conditions of a thermodynamic treatment and also enlightens why the nonadditive composition rule eq. \re{TSALLIS_COMBO} is leading order in a mathematical sense. However, investigating the most researched particular case of nonadditive entropies one can find dozens of different formulas that fulfill either the same, or different but sufficiently simple nonadditive composition rules \cite{Tan01b,Zar10b}. They frequently lead to the same equilibrium distributions in a suitable maximum entropy framework. Moreover, accepting one of the suggested formulas one should ask about the origin of the introduced parameters. E.g. how could we determine the $a=1-q$ parameter of the Tsallis distribution from microscopic models beyond fitting? 

Given a particular microscopic system, such as an ideal gas, a microcanonical treatment assumes that the particles are correlated because the total basic quantities are fixed. A canonical treatment assumes independent distributions with given average basic quantities. The fixed averages are physically provided by suitable reservoirs. A canonical treatment is useful also when the reservoir concept seems purely hypothetic, since its simplicity compared to microcanonical approaches is a great advantage. The Tsallis distribution introduces a particular correlation. It is known that some simple microcanonical distributions are of Tsallis type \cite{Nau11b}. If the origin of Tsallis distributions is universal, what is the distinctive property of these correlations? 

We seek for a generalized entropy formula first examining the two-body thermodynamics of a subsystem and a reservoir, slightly generalizing the results of \cite{BirEta12m}. We consider finite reservoirs and the conservation of the basic quantities: 
$$
X_0^i = X_1^i+X_2^i=const., \quad i=1,...,n.
$$
We aim at a monotonic function of the Boltzmann\,--\,Gibbs entropy $K(S)$ that is maximal in the body-reservoir system and absorbs finite size correlation effects in the Taylor-expansion of the maximum entropy principle. The respective entropy contributions are in general nonadditive and satisfy a composition rule formulated in terms of formal logarithms. We consider homogeneous rules
\be
K(S_{12}) =  K(S_1)+K(S_2).
\ee{ent_comp}

The maximum entropy principle together with the conservation of the basic quantities reads:
\be
K(S(X_1^1,...,X_1^n))+K(S(X_0^1-X_1^1,...,X_0^n-X_1^n))= max.
\ee{max_ent}

The necessary condition of the maximum in case of twice differentiable functions requires vanishing derivatives by the subsystem basic quantities $X^i_1$:
\ba
Y^i_1 &=& K'(S(X_1^1,...,X_1^n))\pd{S}{X_1^i}(X_1^1,...,X_1^n) = \nonumber\\
&=& K'(S(X_0^1-X_1^1,...,X_0^n-X_1^n))\pd{S}{X_1^i}(X_0^1-X_1^1,...,X_0^n-X_1^n) = Y^i_2.
\ea{int_cond}
Subsystem $1$ is much smaller than the reservoir, $X^i_0\gg X^i_1$. This equality in a traditional canonical approach declare the equality of the reservoir and subsystem intensives in the $X^i_1 \rightarrow 0$ limit. Now, considering effects higher order in $X^i_1/X^i_0$ we request that their leading term vanishes on the right hand side of eq. \re{int_cond}. The Taylor series expansion up to the first order reads as
\be
Y^i_1 = K'(S_0)\pd{S_0}{X^i} -\left(K''(S_0)\pd{S_0}{X^i}\pd{S_0}{X^j} + 
  K'(S_0)\pd{^2 S_0}{X^iX^j}\right)dX^j+...
\ee{Taylor_first}
where $S_0 = S(X_0^1,...,X_0^n)$ denotes the function values at the constant basic quantities of the system. It is not obvious that a single function $K(S_0)$ could annul the bracket coefficient of $dX^j$ in general. One of the reasons is that $\pd{S_0}{X^i}\pd{S_0}{X^j}$ is not invertible. Therefore we restrict ourselves to one additional basic quantity that we denote simply by $X$ and the derivative of the entropy by $X$ is denoted b $S'$. Then the previous equality reduces to
\be
Y = K'(S_0)S_0' -\left(K''(S_0)S_0'^2 + K'(S_0)S_0''\right)dX+...
\ee{Taylor_second}
In this case the vanishing linear term requires for a general $K(S)$ that:
\be
\frac{K''(S)}{K'(S)} = -\frac{S''(X)}{S'(X)^2}.
\ee{main_cond}

Here the right hand side is a function of $X$, therefore it is a constant  solving the above equation for an $S$ dependent $K$. Therefore 
\be
\frac{K''(S)}{K'(S)} = a,
\ee{main_ceq}
and the solution of the above equation with the compatibility and triviality conditions $K'(0)=1$ and $K(0)=0$ gives 
\be
K(S) = \frac{e^{aS}-1}{a}.
\ee{Tsallisflog}
The derivatives of the $S(X)$ equation of state do have a physical meaning:
\be
a = 1/\chi(X_0),
\ee{heatcap}
where $\chi$ characterizes the change of the reservoir entropy per unit change of the intensive $Y = S'$. $\chi$ is the {\em generalized susceptibility}. It is remarkable that function \re{Tsallisflog} is identical to the inverse of the formal logarithm of the Tsallis additivity composition \re{TSALLIS_FORMAL_LOG}.

Now we generalize the result of the two body analysis to the classical entropy formula $S = -\sum_i p_i \ln p_i$ and recover the Tsallis entropy formula  as follows:
\be
S_{Tsallis} = \sum_i p_i K(-\ln p_i) = \frac{1}{1-q}\sum_i(p_i^q -p_i),
\ee{Tsallis}
where we have introduced the usual $q=1-a$ notation. This can be considered as a modification of the entropy density in the total average, the entropy density representing the limit of the minimal subsystem in an extended state space. According to the previous section the zeroth law  requires, that the thermodynamic total entropy of the system must be the R\'enyi entropy 
\be
 S_{{\rm R\acute{e}nyi}} = K^{-1}(S_{{\rm Tsallis}}) = \frac{1}{1-q} \ln \sum_i p_i^q. 
\ee{RENYI_AS_FINAL_CONCLUSION}

The corresponding maximum entropy principle reads as 
\be
  \frac{1}{1-q} \ln \sum_i p_i^q  - \beta \sum_i p_i E_i  - \alpha \sum_i p_i  = {{\rm max.}}.
\ee{Renyi_maxent}

This is the basis of calculating the equilibrium distribution in a canonical form.

\section{Summary}

We have derived the requirements of the zeroth law when the basic thermodynamic quantities are nonadditive. We have obtained that then the concept of separate thermodynamic bodies enforces the existence of formal logarithms. The transitivity condition ensure that the formal logarithms represent material  properties. 

Then we were looking for a universal origin of nonadditive compositions and derived the Tsallis entropy formula. We required that the nonadditive composition cancels linear corrections due to a finite $X_0$ basic quantity of the reservoir. That requirement is the principle of Universal Thermostat Independence \cite{BirEta12m}.  This derivation explains the particular functional form of the Tsallis and R\'enyi formulas as generalized entropy expressions. 
 
\ack
The work was supported by the Hungarian National Science Research Fund OTKA NK778816,NK106119,H07/C 74164,K68108, K81161, K104260, NIH TET-10-2011-0061 and ZA-15/2009. GGB also thanks the J\'anos Bolyai Research Scholarship of the HAS.
\vskip 0.5cm

\bibliographystyle{iopart-num}
\providecommand{\newblock}{}

\end{document}